\documentclass[3p,times,procedia]{elsarticle}
\usepackage{nupha_ecrc}

\volume{00}

\firstpage{1}

\journalname{Nuclear Physics A}

\runauth{}

\jid{nupha}

\jnltitlelogo{Nuclear Physics A}

\usepackage{amssymb}
\usepackage{amsthm}
\usepackage{amsmath, bm}

\usepackage[figuresright]{rotating}

\begin{document}

\begin{frontmatter}



%% Instructions from Editor: Please use the following \dochead only in the preprint version (e-print arXiv etc.); 
%% use empty \dochead{} when submitting to Nuclear Physics A!
\dochead{XXVIIIth International Conference on Ultrarelativistic Nucleus-Nucleus Collisions\\ (Quark Matter 2019)}
%% e.g. \dochead{17th International Conference on Dynamical Processes in Excited States of Solids}

\title{Approach to thermalization and hydrodynamics}


\author{Yukinao Akamatsu}

\address{Department of Physics, Osaka University, Toyonaka, Osaka, 560-0043, Japan}

\begin{abstract}
I review recent progress in thermalization in heavy-ion collisions, with particular emphasis on hydrodynamic attractor, and also report recent progress in hydrodynamic fluctuations.
\end{abstract}


\end{frontmatter}

% \linenumbers

\section{Thermalization in heavy-ion collisions}
One of the fundamental questions in relativistic heavy-ion collisions is the early thermalization problem.
Although the initial condition for the relativistic heavy-ion collisions is not fully understood yet, that in the high energy limit is given by classical gluon fields of the coming two nuclei based on the color glass condensate picture \cite{Lappi:2006fp}.
In this picture, strong color electric and magnetic fields are created in the longitudinal direction, whose stress tensor is highly anisotropic, i.e. longitudinal component $T_{\eta\eta}\equiv P_L/\tau^2$ is negative and transverse components $T_{xx}=T_{yy}\equiv P_T$ are positive.
Phenomenological success of hydrodynamic description for collective expansions in the heavy-ion collisions suggests the formation of locally equilibrated isotropic matter as early as 1 fm/$c$.
So far, no theoretical calculation has succeeded in explaining such a rapid transition from strong color field configuration to thermalized system of quarks and gluons, namely the quark-gluon plasma (QGP).

There have been a number of studies of isotropization in the Bjorken expansion, which assumes the boost invariance in the longitudinal direction and translational invariance in the transverse plane.
Such studies range from kinetic theory simulation for weakly coupled quark-gluon plasma to holographic calculation for strongly coupled gauge theory plasma, which contain theoretical descriptions beyond hydrodynamics.
It is found that starting from various initial conditions, evolution of pressure anisotropy $P_L/P_T$ is accurately described by hydrodynamic gradient expansion for $P_L/P_T\gtrsim 0.4$ \cite{Heller:2011ju, Kurkela:2018vqr}.
Since the applicability of gradient expansion is a key assumption for the hydrodynamic description, this surprising results indicate that early thermalization or isotropization may not be necessary in the heavy-ion collisions but at the same time raise a following question: how far from equilibrium is hydrodynamics applicable?

Conventionally, hydrodynamics is regarded as a low energy effective theory near thermal equilibrium.
Its equation of motion is written solely by hydrodynamic modes such as conserved densities, Nambu-Goldstone modes, and U(1) gauge fields of unbroken gauge symmetries, using the gradient expansion and the symmetry principles.
In addition to these hydrodynamic modes, the system also contains non-hydrodynamic modes that evolve fast.
When the spatial variation is small, the non-hydrodynamic modes can be quickly adjusted to the surrounding environment so that a few local gradients can characterize their response to the variation.
When the spatial variation is large, it usually takes time even for the non-hydrodynamic modes to get adjusted to the environment.
However, once they are adjusted to the large gradient, they are not dynamical anymore and the number of effective degrees of freedom is reduced.
Therefore, once the non-hydrodynamic modes are ineffective and if their non-perturbative response to the large gradient is known, hydrodynamic description (only in terms of the hydrodynamic modes) can be extended further out of equilibrium \cite{Romatschke:2017vte}.
In the following section, I will review hydrodynamic attractor which explicitly demonstrates how far-from-equilibrium hydrodynamics can be constructed.
Then, I will also review hydrodynamic fluctuations which are caused by the equilibrated non-hydrodynamic modes and
become important when the number of particles is small.

\section{Hydrodynamic attractor}
\subsection{Physical mechanism of hydrodynamic attractor}
\begin{figure}
\centering
\vspace{0cm}
\includegraphics[angle=0, width=0.37\linewidth]{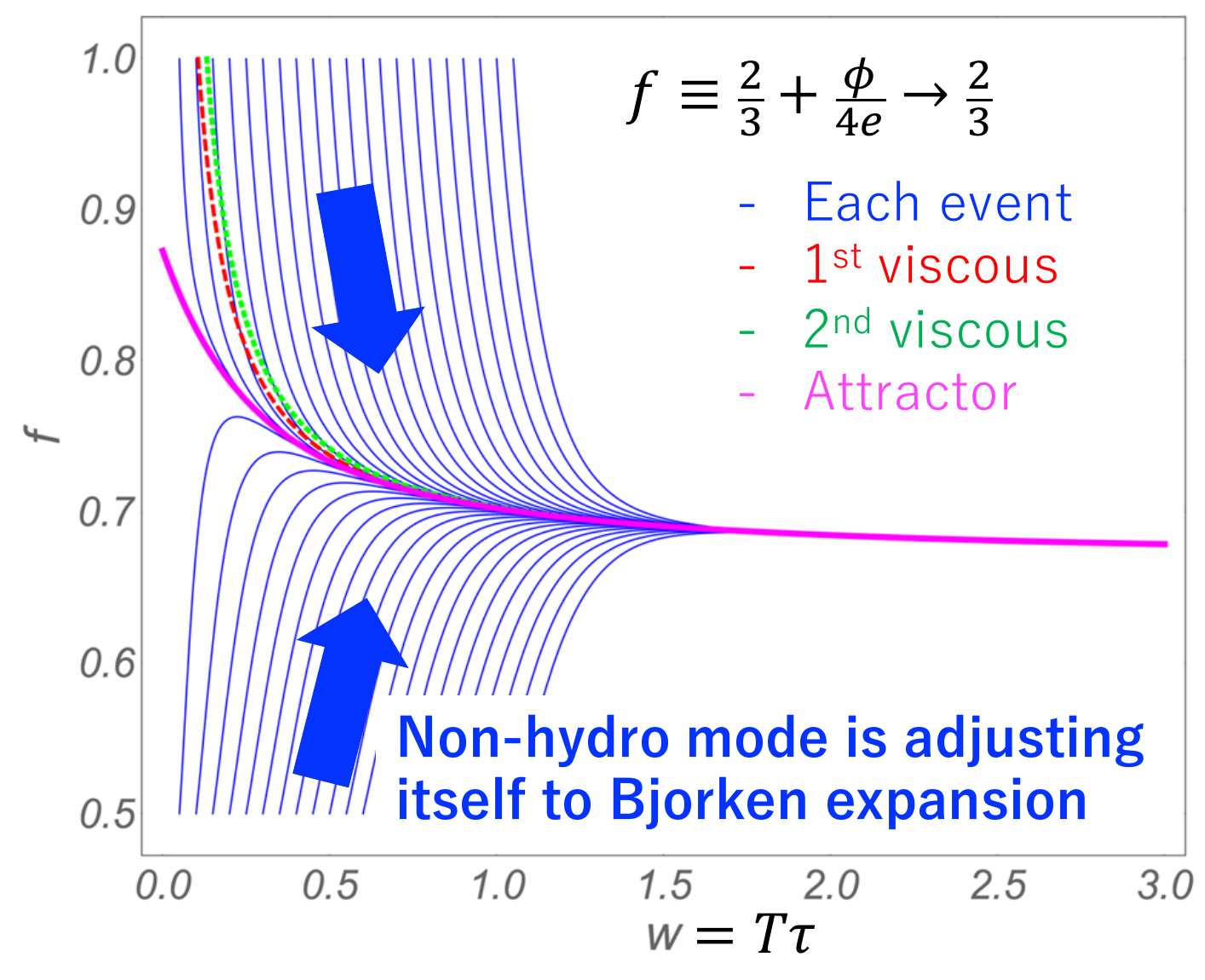}
\vspace{0cm}
\includegraphics[angle=0, width=0.4\linewidth]{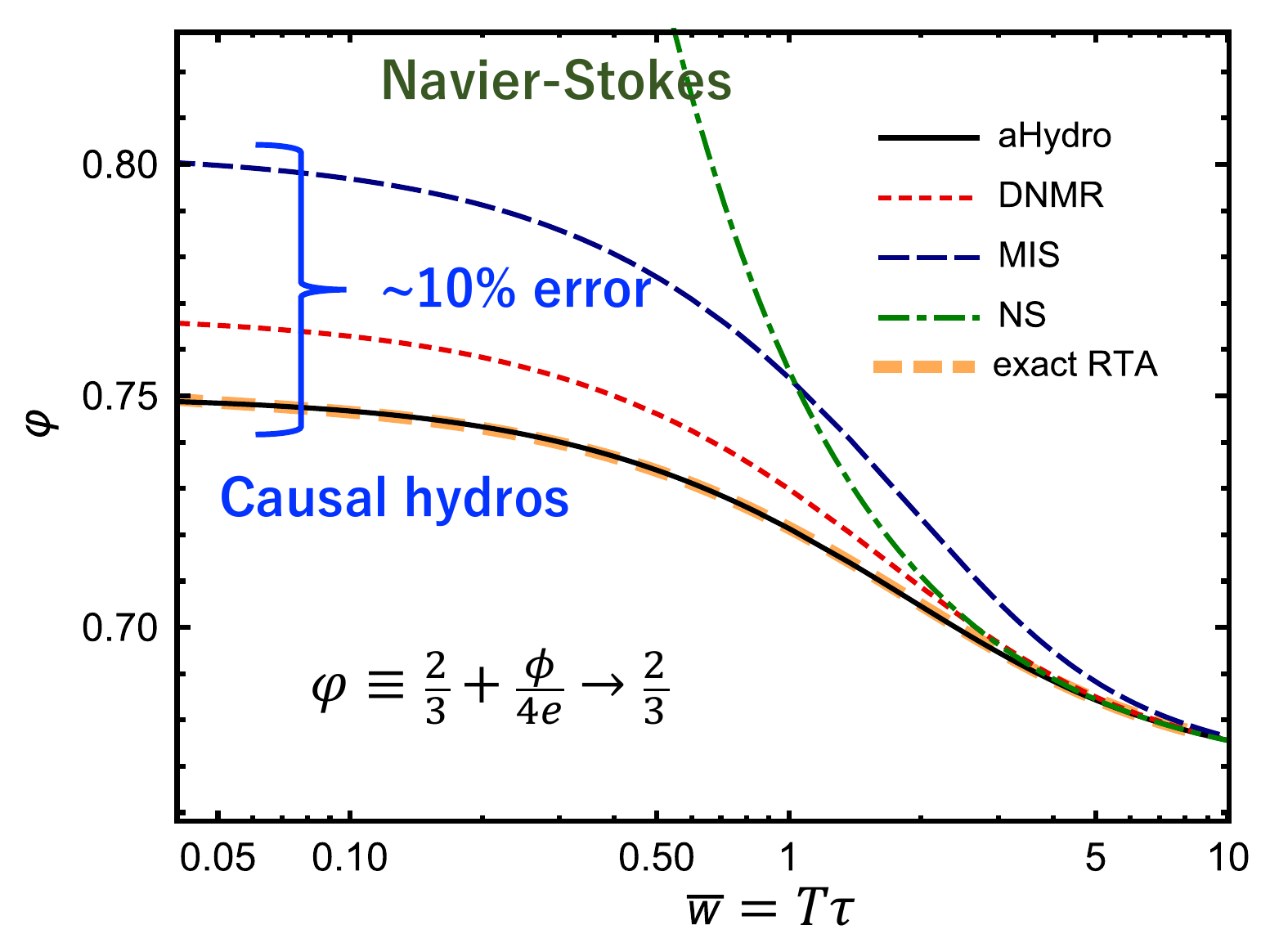}
\caption{(Left) Time evolutions of pressure anisotropy of conformal causal hydrodynamics from different initial conditions in the Bjorken expansion.
Figure adapted from \cite{Heller:2015dha}.
(Right) Hydrodynamic attractors of pressure anisotropy for conformal kinetic theory with relaxation time approximation (RTA) and its various approximations in the Bjorken expansion.
Figure adapted from \cite{Strickland:2017kux}.
}
\label{fig:Attractors}
\end{figure}

Let us start by an example of the hydrodynamic attractor for conformal causal hydrodynamics in the Bjorken expansion \cite{Heller:2015dha}.
In this case, the hydrodynamic mode is the energy density $e$ and non-hydrodynamic mode is the shear mode $\phi$ and they evolve according to 
\begin{align}
-\frac{de}{d\tau}=\frac{e+P(e)-\phi}{\tau}, \quad
-\tau_{\pi}\frac{d\phi}{d\tau}=\phi-\frac{4\eta}{3\tau} + \frac{4\tau_{\pi}\phi}{3\tau}+\frac{\lambda_1\phi^2}{2\eta^2}.
\end{align}
The energy density gets smaller by the Bjorken expansion and the longitudinal work.
The shear mode relaxes toward the Navier-Stokes value $4\eta/3\tau$, but the Bjorken expansion ($4\tau_{\pi}\phi/3\tau$) disturbs the relaxation.
There is also a nonlinear self interaction term $\lambda_1\phi^2/2\eta^2$ due to the conformal symmetry.
The left panel of Fig.~\ref{fig:Attractors} shows time evolution of $f\equiv \frac{2}{3} + \frac{\phi}{4e}$ which approaches 2/3 in the long time.
Each blue line represents the evolution from different initial conditions and the non-hydrodynamic mode is getting adjusted to the Bjorken expansion. 
All the curves eventually asymptote to the attractor solution.
It is clear that the attractor can characterize the solutions beyond the regime where the gradient expansions work.

The right panel of Fig.~\ref{fig:Attractors} compares hydrodynamic attractors in the Bjorken expansion for conformal kinetic theory with relaxation time approximation (RTA) and its various approximations \cite{Strickland:2017kux}.
The RTA kinetic theory in this setup solves
\begin{align}
\left[\partial_{\tau} - \frac{p_z}{\tau}\partial_{p_z}\right]f(\bm p, \tau)=-\frac{f(\bm p, \tau)-f_{\rm eq}(\bm p;T_{\rm eff})}{\tau_R}, \quad
\tau_R(\tau)\propto T_{\rm eff}(\tau)^{-1}\propto e(\tau)^{-1/4},
\end{align}
where $f(\bm p,\tau)$ is the momentum distribution in the local rest frame.
It contains one hydrodynamic mode, namely the energy density, and infinitely many non-hydrodynamic modes from the higher moments of $f(\bm p, \tau)$.
All the approximations here result in different versions of causal hydrodynamics, which contain single non-hydrodynamic mode in this setup.
In spite of such a big difference in the non-hydrodynamic sector, the hydrodynamic attractors of causal hydrodynamics are within 10\% error from that of the RTA kinetic theory as can be seen in the figure.

\begin{figure}
\centering
\vspace{0cm}
\includegraphics[angle=0, width=0.8\linewidth]{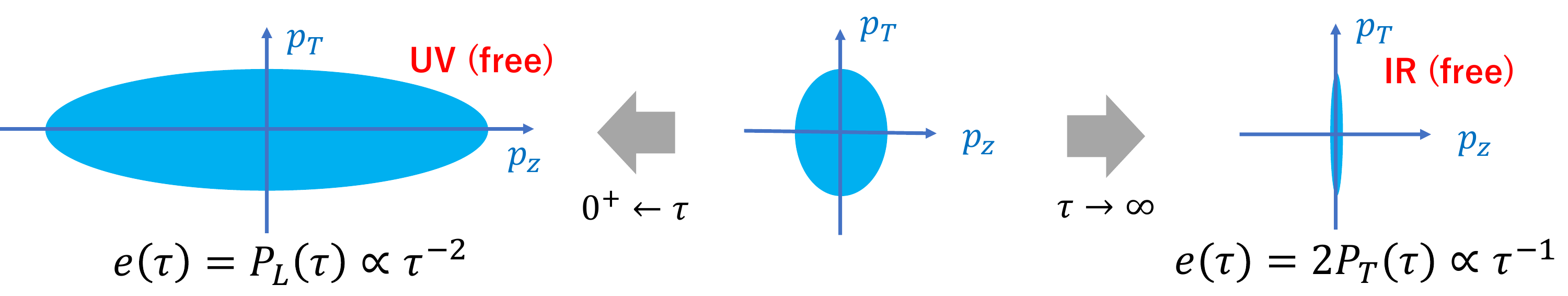}\\
\hspace{5.8cm}
\includegraphics[angle=0, width=0.44\linewidth]{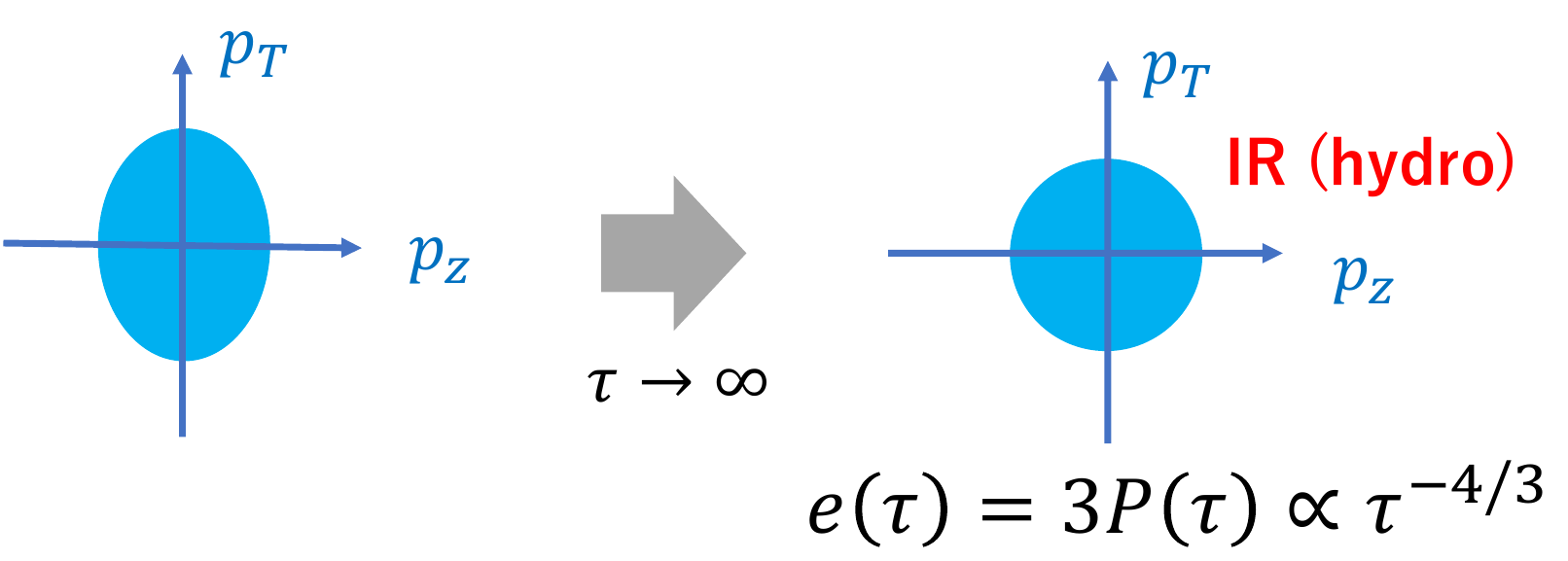}
\caption{(Upper) Asymptotic momentum distributions of free streaming particles in the Bjorken expansion.
(Lower) Asymptotic momentum distribution of interacting particles in the Bjorken expansion.
}
\label{fig:FPs}
\end{figure}

The reason why the causal hydrodynamics is a good approximation to the RTA kinetic theory is given by asymptotic behaviors of both theories \cite{Blaizot:2017ucy, Blaizot:2019scw}.
Let us first briefly analyze the asymptotic behavior of the RTA kinetic theory.
At initial times $\tau\ll \tau_R$, the collision term can be neglected because almost no collision takes place and the RTA kinetic theory describes the free streaming particles.
In this free streaming limit, typical longitudinal momentum $p_z$ gets smaller and smaller because a particle with finite momentum moves away from the rest frame.
Therefore the momentum distribution becomes flat and the longitudinal pressure vanishes $P_L\to 0$ as shown in the upper panel of the Fig.~\ref{fig:FPs}.
The long-time behavior of the energy density is thus $e(\tau)\propto\tau^{-1}$.
Going backward in time ignoring the physical inapplicability of the kinetic theory for a moment, the longitudinal pressure becomes larger and larger $P_L\simeq e$ and the early-time behavior of the energy density is $e(\tau)\propto\tau^{-2}$.
When the collision term is included for $\tau\gtrsim\tau_R$, the long-time asymptotic behavior is the well-known hydrodynamic scaling and the energy density is $e(\tau)\propto\tau^{-4/3}$.
In this way, it is the exponents of physical quantities that characterize the asymptotic behaviors of different regimes.

At this point, it is useful to perform fixed point analysis for the logarithmic growth rates of physical quantities.
Let us define $g_n$ as the logarithmic growth rates of the $n$-th moments of the momentum distribution $\mathcal L_n\equiv\int \frac{d^3p}{(2\pi)^3p_0} |\bm p|^2 P_{2n}(p_z/|\bm p |)f(\bm p, \tau)$, where $P_{2n}$ is a Legendre polynomial of order $2n$.
The first two are:
\begin{align}
g_0(\tau)\equiv \frac{d\log\mathcal L_0}{d\log\tau}=\frac{d\log e}{d\log\tau}, \quad
g_1(\tau)\equiv \frac{d\log\mathcal L_1}{d\log\tau}=\frac{d\log (P_L-P_T)}{d\log\tau}.
\end{align}
Each solution of the RTA kinetic theory can be mapped to a trajectory in the space of $g_n \ (n=0, 1, 2, \cdots)$.
As shown above for $g_0$, there are three fixed points in the RTA kinetic theory.
With a more elaborate analysis, one can obtain these fixed points in the space of $g_n$:
(i) free-streaming UV fixed point ($g_n=-2$),
(ii) free-streaming IR fixed point ($g_n=-1$), and
(iii) hydrodynamic IR fixed point ($g_n=-(4+2n)/3$)
\footnote{
UV (ultraviolet) stands for the early-time behavior and IR (infrared) stands for the long-time behavior.
Note that free-streaming IR fixed point is not a true fixed point in the full RTA kinetic theory.
}.
The system starts off somewhere and approaches the free-streaming IR fixed point and then dragged to the hydrodynamic IR fixed point.
The attractor is a solution that starts from the free-streaming IR fixed point and evolves toward the hydrodynamic IR fixed point.
Note that the approach to the free-streaming IR fixed point is power-law like without any intrinsic time scale while that to the hydrodynamic IR fixed point takes place with a relaxation time scale \cite{Kurkela:2019set}.

We can perform a similar analysis to show that the causal hydrodynamics, which corresponds to 2-moment truncation of the RTA kinetic theory, has corresponding fixed points close to those in the RTA kinetic theory.
Truncated equations of motion are written in the matrix form
\begin{align}
\tau\frac{d}{d\tau}\begin{pmatrix}
\mathcal L_0\\
\mathcal L_1
\end{pmatrix}
=-\begin{pmatrix}
4/3 & 2/3 \\
8/15 & 38/21
\end{pmatrix}
\begin{pmatrix}
\mathcal L_0\\
\mathcal L_1
\end{pmatrix}
-\begin{pmatrix}
0\\
\frac{\tau}{\tau_R}\mathcal L_1
\end{pmatrix},
\end{align}
and one can readily find the free-streaming UV and IR fixed points at $(g_0, g_1)\approx(-2.21, -2.21)$ and $(g_0, g_1)\approx(-0.93, -0.93)$ and the hydrodynamic IR point at $(g_0, g_1)=(-4/3, -2)$, which are close to or the same with the fixed points of the RTA kinetic theory.
This means that the causal hydrodynamics captures the global features of the RTA kinetic theory solutions even at early times and the hydrodynamic attractor of the latter is well approximated by that of the former.

\subsection{New developments in the attractor}
Far-from-equilibrium hydrodynamics based on the hydrodynamic attractor is a very rapidly developing field \cite{Heller:2011ju, Romatschke:2017vte, Heller:2015dha, Strickland:2017kux, Lublinsky:2007mm, Heller:2013fn, Romatschke:2016hle, Behtash:2017wqg, Heller:2016rtz, Strickland:2018ayk, Behtash:2018moe, Behtash:2019txb, Denicol:2019lio, Jaiswal:2019cju, Behtash:2019qtk, Chattopadhyay:2019jqj}.
Theoretical methods include Borel resummation of large order gradient expansion, slow-roll approximation, and direct numerical simulations.
The fixed point analysis introduced above not only demonstrates a useful approximation scheme, which is the moment truncation for the RTA kinetic theory, but also points out that the existence of the free-streaming IR fixed point is essential to the hydrodynamic attractor \footnote{
The AdS/CFT calculation for strongly coupled plasma in the Bjorken expansion indicates that there is no early-time IR fixed point and the attractor behavior emerges only at late time about the relaxation time scale \cite{Kurkela:2019set}.
This is in sharp contrast to the weakly coupled plasma, where free-streaming behavior is naturally expected.
}.
There are also a few interesting developments reported in this conference, which I briefly summarize here.

The fixed point analysis explained above is for the RTA kinetic theory, which is governed by linearized Boltzmann equation.
The QCD effective kinetic theory solves full collision term of 1-to-2 and 2-to-2 processes for hard gluons and quarks.
This theory is known to contain a non-thermal fixed point far from equilibrium, namely overpopulated anisotropic plasma, which is the first stage of the bottom-up thermalization scenario \cite{Baier:2000sb}.
Numerical analysis of the QCD effective kinetic theory suggests that the momentum distribution function exhibits scaling behavior $f_g(p_{\perp}, p_z, \tau)=\tau^{\alpha(\tau)}f_S(\tau^{\beta(\tau)}p_{\perp}, \tau^{\gamma(\tau)}p_z)$ even before the system reaches the non-thermal fixed point \cite{Mazeliauskas:2018yef}.
The time-dependent exponents $\alpha(\tau)$, $\beta(\tau)$, and $\gamma(\tau)$ slowly approach the values $\alpha\approx -0.7$, $\beta=0$, and $\gamma\approx 0.3$ at the non-thermal fixed point \cite{Baier:2000sb, Bodeker:2005nv}.
This phenomenon is called {\it pre-scaling}.
In the pre-scaling regime, the momentum distribution is characterized by only three scaling exponents which evolve slowly in time, suggesting an attractor behavior.

Another interesting development is a new concept of {\it adiabatic hydrodynamics} \cite{Brewer:2019oha}.
The adiabatic hydrodynamics is a way to extract hydrodynamic attractor from a microscopic theory, such as kinetic theory, by tracing the slowest configuration at each time, assuming all the other configurations decay quickly.
This idea is borrowed from a well-known adiabatic approximation of quantum mechanics with an energy gap.
The adiabatic hydrodynamics is applied to the RTA kinetic theory, whose linearized Boltzmann equation finds a clear correspondence to the time-dependent Schr\"odinger equation, and qualitatively reproduced the attractor of the RTA kinetic theory.
It suggests that the adiabatic ``ground state" effectively selects the attractor solution.

\subsection{Phenomenological application of the attractor}
\begin{figure}
\centering
\vspace{0cm}
\includegraphics[angle=0, width=0.35\linewidth]{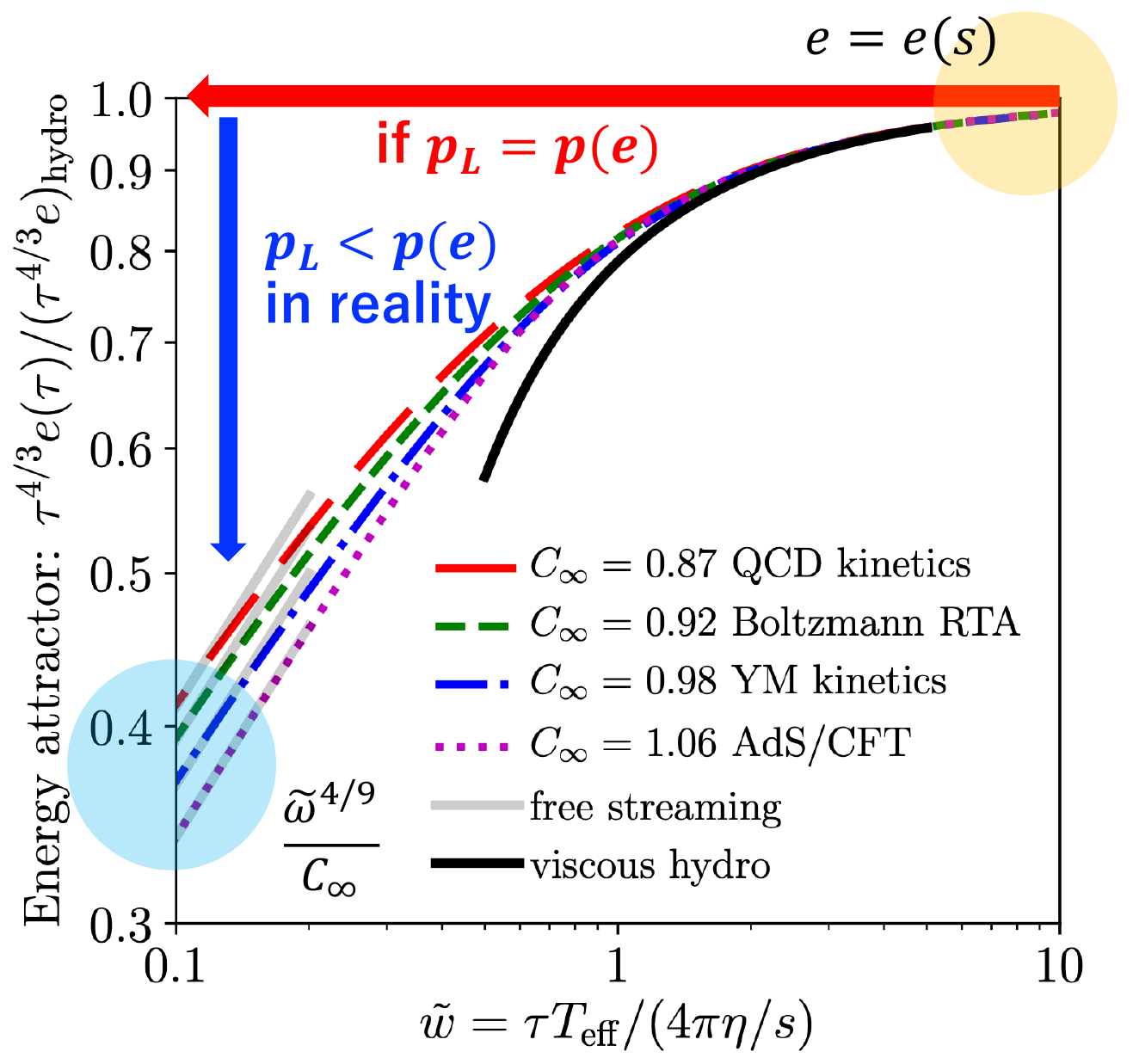}
\includegraphics[angle=0, width=0.50\linewidth]{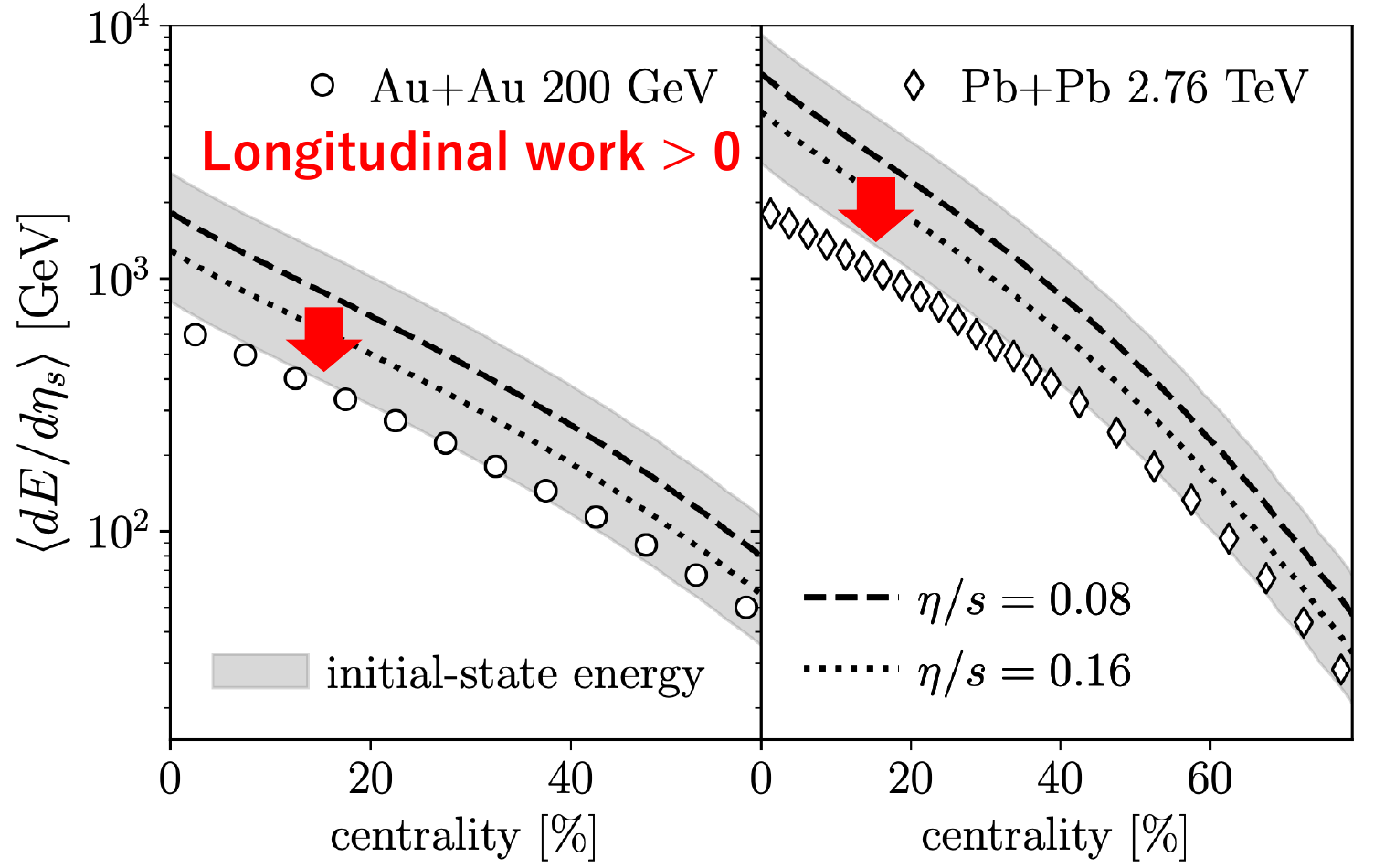}
\caption{(Left) Energy attractors of different theories in the Bjorken expansion.
Both the energy density and the time scale is properly normalized so that the attractors of different theories behave similarly.
(Right) Initial energy per spacetime rapidity $\langle dE/d\eta_s\rangle$ is estimated by the formula \eqref{eq:initial_E_formula}.
Measured energy per particle rapidity $dE/dy$ is compared.
Both figures adapted from \cite{Giacalone:2019ldn}.
}
\label{fig:EnergyAttractor}
\end{figure}

Once the hydrodynamic attractor is known, is there anything interesting we can discuss with it? 
There are actually very interesting phenomenological applications of the hydrodynamic attractor \cite{Giacalone:2019ldn, Kurkela:2018xxd, Kurkela:2018oqw, Schlichting:2019abc}, from which I pick one recent example \cite{Giacalone:2019ldn}.
The left panel of Fig.~\ref{fig:EnergyAttractor} shows energy attractors calculated by several different models ranging from weakly coupled QCD plasma to supersymmetric Yang-Mills theory at strong coupling.
The energy density is normalized to the ideal hydrodynamic solution $\tau^{4/3}e(\tau)/(\tau^{4/3}e)_{\rm hydro}$ so that it approaches 1 at late times.
Going backward in time, the longitudinal pressure is weaker than that of the ideal hydrodynamics because the system is still anisotropic.
Therefore the normalized energy density gets smaller at earlier times.
Important observation is that all the attractors of different theories with different time scales behave similarly once the time is scaled by the equilibrium relaxation time $\tilde\omega =\tau T_{\rm eff}/(4\pi\eta/s)$.
Independently of the model details, we can compare the initial energy density to the late time energy density and late time entropy density via the equation of state $e=(\nu_{\rm eff}\pi^2/30)T^4=(3/4)sT$:
\begin{align}
\tau_{\rm ini}^{4/3}e(\tau_{\rm ini})\approx
\tau_{\rm hydro}^{4/3}e(\tau_{\rm hydro}) \tilde\omega^{4/9}, \quad
\tau_{\rm hydro}s(\tau_{\rm hydro})\approx
\left(4\pi\eta/s\right)^{1/3}\nu_{\rm eff}^{1/3} e(\tau_{\rm ini})\tau_{\rm ini}.
\end{align}
The late time entropy only increases slightly by viscous effects even after the transverse expansion sets in so that it is almost conserved until the freezeout.
Therefore a dominant part of the entropy production is related to the initial energy density by using the attractor.

The formulae can be used inversely to estimate the initial energy density from the charged particle multiplicity:
\begin{align}
\label{eq:initial_E_formula}
\tau_{\rm ini}e(\tau_{\rm ini})
\sim 30\left(\frac{4\pi\eta}{s}\right)^{-1/2}\nu_{\rm eff}^{-1/2}
\left(\frac{dN_{\rm ch}}{A_Td\eta}\right)^{3/2},
\end{align}
where $A_T$ is the transverse area of the fireball.
The right panel of Fig.~\ref{fig:EnergyAttractor} compares the measured energy per particle rapidity $dE/dy$ and the initial energy per spacetime rapidity $\langle dE/d\eta_s\rangle$ estimated by the above formula and the difference must come from the amount of work done by the thermalizing QGP.
In the estimate of the initial energy density, typical values are adopted for the parameters such as shear viscosity to entropy density ratio.
Since the longitudinal work must be positive, some larger values for the ratio of shear viscosity to entropy density is rejected only by using independent experimental data.

\section{Hydrodynamic fluctuations}
\subsection{Kinetic regime of hydrodynamic fluctuations and hydro-kinetic theory}
After enough time has passed, the non-hydrodynamic modes are almost equilibrated and their effects can be described by linear response and noise, which are related by the fluctuation-dissipation theorem \cite{Kapusta:2011gt}.
The hydrodynamic equation is still the conservation law but the energy-momentum tensor contains noise contributions.
In the presence of noise, hydrodynamic fluctuations are excited, evolve, and relax on the background flow.
In contrast to the non-hydrodynamic modes, some hydrodynamic fluctuations can never get equilibrated on the background flow.
The kinetic regime $k_*$ characterizes the wave number at which the time scales of relaxation $1/\gamma_{\eta}k^2=1/(\eta/sT)k^2$ and background expansion $\tau$ balance.
The hydro-kinetic theory is the theory that describes the dynamics of particle-like modes in the kinetic regime $k_*\sim 1/\sqrt{\gamma_{\eta}\tau} \gg 1/\tau$, which are the most important out-of-equilibrium fluctuations \cite{Akamatsu:2016llw, Akamatsu:2017rdu, Martinez:2018wia}.

The hydrodynamic fluctuations in the kinetic regime give two kinds of nonlinear contributions to the total energy-momentum tensor.
Their contribution is sensitive to the hard cutoff $\Lambda$ of the wavenumber and the cutoff dependence needs to be canceled by renormalizing the background energy density $e_0$, pressure $p_0$ and viscosities $\eta_0, \zeta_0$ \cite{Akamatsu:2016llw, Akamatsu:2017rdu, Kovtun:2011np}:
\begin{align}
e(T) &= e_0(T;\Lambda) + \frac{T\Lambda^3}{2\pi^2}, \quad
p(T) = p_0(T;\Lambda) + \left(1+\frac{T}{2}\frac{dc_{s0}^2}{dT}\right) \frac{T\Lambda^3}{6\pi^2}, \\
\eta(T) &=
\eta_0(T;\Lambda) 
+\frac{T\Lambda}{30\pi^2}
\left[
\frac{e_0+p_0}{\zeta_0 + \frac{4}{3}\eta_0}
+\frac{7(e_0+p_0)}{2\eta_0}
\right], \\
\zeta(T)&=
\zeta_0(T;\Lambda)+\frac{T\Lambda}{18\pi^2}
\left[
\begin{aligned}
\left(1+\frac{3T}{2}\frac{dc_{s0}^2}{dT} -3c_{s0}^2\right)^2\frac{e_0 + p_0}{\zeta_0+\frac{4}{3}\eta_0} 
+4\left(1-3c_{s0}^2\right)^2\frac{e_0 + p_0}{2\eta_0}
\end{aligned}
\right].
\end{align}
This means that the fluctuating hydrodynamics is a cutoff dependent theory like quantum/statistical field theories.
Physically, the energy density $e_0(\Lambda)$, pressure $p_0(\Lambda)$, and viscosities $\eta_0(\Lambda), \zeta_0(\Lambda)$ in the equations of fluctuating hydrodynamics represent the equilibrium properties of the modes above the cutoff $\Lambda$ including the hydrodynamic fluctuations with $k\geq\Lambda$.
After the renormalization, the remaining contribution to the stress tensor from the hydrodynamic fluctuations $[T_{ij}]_{\rm fluct}$ is genuinely out-of-equilibrium.
It is estimated by counting the number of modes $\int_{k_*}1\sim k_*^3$ and each mode contributes by $\sim T$, yielding $[T_{ij}]_{\rm fluct}\sim Tk_*^3\sim T/(\gamma_{\eta}\tau)^{3/2}$.
In the Bjorken expansion, the noise averaged longitudinal pressure is given by \cite{Akamatsu:2016llw}
\begin{align}
\left\langle \tau^2T^{\eta\eta}\right\rangle = p - \frac{4\eta}{3\tau}
+ \frac{1.08318T}{(4\pi \gamma_\eta \tau)^{3/2}}
+ \cdots.
\end{align}
In contrast to the non-hydrodynamic modes which decay with a finite relaxation time, the out-of-equilibrium contribution $[T_{ij}]_{\rm fluct}$ from hydrodynamic fluctuations lasts long (the long-time tail \cite{Kovtun:2003vj}).
It gets smaller because the phase space volume of the kinetic regime $k_*^3$ decreases by the Bjorken expansion.

Recently, the hydro-kinetic theory for a general background flow is developed \cite{An:2019osr}.
Since the hydrodynamic fluctuations in the kinetic regime is a particle-like mode on the background flow, this generalization allows for all kinds of local flows in the background, namely rotation and acceleration as well as shear and bulk flows.
Also, this generalization explicitly takes account of non-uniform background flow by using the mathematical notions of {\it confluent correlators} and {\it confluent derivatives} so that the local description is obtained in a covariant manner.
For example, equation of motion for phonons on the local rest frame is derived and it contains the Coriolis and inertial forces in the presence of rotation and acceleration in addition to the actual forces by potential and Hubble-like term.
As for the transverse mode, there is no particle interpretation because it does not propagate.
Hydro-kinetic theory on the general background may provide a new method to solve the fluctuating hydrodynamics by solving coupled equations for the background flow and the phonon gas.

\subsection{Numerical simulation of fluctuating hydrodynamics}
There are also direct simulations of the fluctuating hydrodynamics on the numerical grid \cite{Young:2013fka, Young:2014pka, Yan:2015lfa, Chattopadhyay:2017rgh, Singh:2018dpk, Sakai:2019qm}.
One of the important phenomenological consequences of the thermal noise is {\it longitudinal decorrelation} of the particle flows \cite{Sakai:2019qm}.
The longitudinal correlation of particle flows with $\eta_a$ and $-\eta_a$ can be measured by $r_n(\eta_a,\eta_b)\equiv V_{n\Delta}(-\eta_a, \eta_b)/V_{n\Delta}(\eta_a, \eta_b)$ with $V_{n\Delta}(\eta_a, \eta_b)\equiv \langle\cos(n\Delta\phi)\rangle_{\eta_a,\eta_b}$, where $\eta_b$ is a reference pseudorapidity at some forward region $3<\eta_b <4$.
From the simulation, the effects of initial and thermal fluctuations are found to be comparable, indicating that it is essential to consider the effect of thermal fluctuations in order to study the initial longitudinal fluctuations quantitatively.

In numerical simulations of the relativistic hydrodynamics, causal hydrodynamics is implemented in order to avoid superluminal propagation and to achieve numerical stability.
The fluctuating hydrodynamics can also be extended to the causal fluctuating hydrodynamics by introducing a noise term to the constitutive equation.
Schematically, the constitutive equation and the noise correlation take the following form \cite{Murase:2019cwc}
\begin{align}
(1+\tau_R D)\pi = \pi_{\rm NS} + \xi, \quad
\langle\xi(x)\xi(x')\rangle = T\kappa\delta(x-x')\left[
2+\tau_R D\ln(T\kappa/\tau_R) - \tau_R\theta
\right],
\end{align}
where $D\equiv u^{\mu}\partial_{\mu}$ denotes the local time derivative, $\theta\equiv\partial_{\mu}u^{\mu}$ is the expansion rate, and $\kappa$ is the transport coefficient.
In the causal hydrodynamics, the memory effect is introduced and it takes finite time $\tau_R$ for $\pi$ to reach the Navier-Stokes value $\pi_{\rm NS}$.
Correspondingly, the noise strength must also take account of the fact that the background is evolving because of the non-instantaneous nature of the theory.
This {\it modified fluctuation-dissipation relation} results in a visible consequence in the distribution of produced entropy.
By applying the fluctuation theorem, one of the fundamental theorems in non-equilibrium physics, to the Bjorken expansion, the average and the variance of the entropy production rate are related \cite{Hirano:2018diu}.
Numerical simulation of the fluctuating causal hydrodynamics shows that the fluctuation-dissipation relation needs to be modified as proposed above.

\section{Summary}
In a simple set up of the RTA kinetic theory in the Bjorken expansion, we understand the global feature of the early time dynamics toward hydrodynamization, hydrodynamic attractor, and far-from-equilibrium hydrodynamics, using the fixed point structure.
Further developments in the hydrodynamic attractor based on the microscopic theory would be desired, such as the role of plasma instability, the effect of 3-dimensional or transverse expansion, and the description of freezeout fixed point which is an inevitable consequence of the fireball.
Experimental data should be interpreted in this whole picture with various possible fixed points (free streaming, non-thermal, hydrodynamic, freezeout, $\cdots$) and we need to recognize which fixed point property we are observing.
Scanning the data by the multiplicity might be an interesting way to explore the different trajectories.
The developments in the hydrodynamic fluctuations would also help improve the theoretical predictions in the smaller collision systems in this challenging program.

\subsection*{Acknowledgments}
The author is supported by JSPS KAKENHI Grant Number JP18K13538 and by Kohsuke Yagi Quark Matter Award 2019.
The author also thanks Masayuki Asakawa, Jean-Paul Blaizot, Jasmine Brewer, Hirotsugu Fujii, Kenji Fukushima, Michal Heller, Tetsufumi Hirano, Masakiyo Kitazawa, Aleksi Kurkela, Aleksas Mazeliauskas, Koichi Murase, Azumi Sakai, Soeren Schlichting, Chun Shen, Derek Teaney, Li Yan, and Yi Yin, for valuable advices, discussions, and encouragements.

\label{}



%% References


\bibliographystyle{elsarticle-num}
\bibliography{qm2019_proc.bib}






% \end{thebibliography}

\end{document}